\documentclass[12pt]{amsart}
\usepackage{amssymb,latexsym}

\title[Rotating traversable wormholes]
{Axially symmetric rotating traversable wormholes}
\author{Peter K. F. Kuhfittig}
\address{Department of Mathematics\\
Milwaukee School of Engineering\\
Milwaukee, Wisconsin 53202-3109}
\date{\today}
\begin{document}

\maketitle 

\begin{abstract}
This paper generalizes the static and spherically symmetric traversable wormhole
geometry to a rotating axially symmetric one with a time-dependent angular velocity
by means of an exact solution.  It was found that the violation of the weak energy
condition, although unavoidable, is considerably less severe than in the static 
spherically symmetric case.  The radial tidal constraint is more easily met due to
the rotation.  Similar improvements are seen in one of the lateral tidal 
constraints.  The magnitude of the angular velocity may have little effect on the
weak energy condition violation for an axially symmetric wormhole.  For a spherically
symmetric one, however, the violation becomes less severe with increasing angular
velocity.  The time rate of change of the angular velocity, on the other hand, was
found to have no effect at all.  Finally, the angular velocity must depend only on
the radial coordinate, confirming an earlier result.
\end{abstract}

PAC number(s): 04.20.Jb, 04.20.Gz

\section{Introduction}
It was recognized by Flamm ~\cite{lF16} in 1916 that our universe may not
be simply connected:  there may exist handles or tunnels, now called
wormholes, in the spacetime topology linking widely separated regions 
of our universe or even connecting us with different universes
altogether.  That such wormholes may be traversable by humanoid
travelers was first conjectured by Morris and Thorne~\cite{MT88},
thereby suggesting that interstellar travel and even time travel may 
some day be possible.  For a detailed discussion see the book by 
Visser~\cite{mV96}.  

Morris-Thorne (MT) wormholes are static and spherically symmetric
and connect asymptotically flat spacetimes, here assumed to be 
isometric.  Adopting units in which $c=G=1$, the metric for this 
wormhole is given by
\begin{equation}\label{E:line1}
    ds^2=-e^{2\Phi(r)}dt^2+\frac{dr^2}{1-b(r)/r}+r^2(d\theta^2
    +\text{sin}^2\theta\,d{\phi}^2);
\end{equation}
$\Phi(r)$ is called the \emph{redshift function} and $b(r)$ the
\emph{shape function}.  The shape function describes the spatial
shape of the wormhole when viewed, for example, in an embedding
diagram, described below.

To hold such a wormhole open, violations of certain energy
conditions proved to be unavoidable.  More precisely, all known forms
of matter obey the weak energy condition (WEC)
$T_{\alpha\beta}\mu^{\alpha}\mu^{\beta}\ge0$ for all timelike
vectors and, by continuity, all null vectors (Friedman~\cite{jF97}).
Matter that violates this condition is called \emph{exotic}
by Morris and Thorne.

Various attempts have been made to generalize the MT wormhole by
giving up spherical symmetry \cite{mV96} or by including 
time-dependence.  A particular interesting example of the latter
is the inclusion of a de Sitter scale factor multiplying the spatial
part of the metric.  The goal was to study the possibility of
enlarging a wormhole pulled out of the spacetime foam to 
macroscopic size (Roman~\cite{tR93}).  A similar scale factor was 
used by Kim~\cite{swK96}.  Yet another possibility is the use of
a conformal factor $\Omega(t)~$\cite{lA98,sK94,KS96,WL95}:
\begin{equation}\label{E:line2}
    ds^2=\Omega(t)\left[-e^{2\Phi(r)}dt^2+e^{2\Lambda(r)}dr^2+
    r^2(d\theta^2+\text{sin}^2\theta\,d{\phi}^2)\right].
\end{equation} 
For a metric with time-dependent functions $\Phi$ and $\Lambda$,
see Ref.~\cite{pK99}.  All these studies include a discussion of 
the WEC violation.  Traversability conditions are investigated in
\cite{lA98} and \cite{pK99}.

In this paper we generalize the MT wormhole in another direction 
by assuming the wormhole to be rotating, not necessarily at a
constant rate, and by dropping the assumption of spherical 
symmetry.  Instead, the wormhole is assumed to be axially
symmetric, i.e., symmetric with respect to the axis of rotation.
Stationary axially symmetric wormholes are discussed by Teo~
\cite{eT98}.  It is shown that the WEC is indeed violated but 
that a traveler would not necessarily come into contact with 
any of the exotic matter.  These wormholes may also have an
ergoregion, where a particle cannot remain stationary with
respect to spatial infinity.  This is an extreme example of the
well-known dragging effect in general relativity.

The main purpose of this paper is to discuss both the energy 
violation and the
traversability conditions by first finding an exact solution.
Possible restrictions on the metric coefficients recently
proposed by Perez Bergliaffa and Hibberd \cite{PB02} are
discussed in Section~\ref{S:additional}.

Proposals to search for naturally occurring wormholes, if they 
exist, go back at least to 1995~\cite{jC95}.  For a summary of
these findings see Ref.~\cite{mS02}.  While definite conclusions 
are still lacking, the possible existence of wormholes or  
the existence of negative mass cannot be ruled out.

\section{The metric}\label{S:metric}
The study of spacetimes that are both stationary and axially 
symmetric has a long history \cite{aP66,bC69,bC87}.  A spacetime 
is stationary if it possesses a time-like Killing vector field
$\eta^a=(\partial/\partial t)^a$ generating invariant time 
translations.  Axially symmetric is formally defined as
possessing a space-like Killing vector field $\xi^a=(\partial
/\partial \phi)^a$ generating invariant rotations with respect 
to $\phi$. 

Suitable metrics for stationary axially symmetric fields are
discussed by Islam~\cite{jI85}.  We will adopt the metric suggested
by Teo~\cite{eT98}, since it appears to be best-suited for obtaining
an exact solution:
\begin{equation}\label{E:line3}
    ds^2=-N^2dt^2+e^{\mu}dr^2+r^2K^2\left[d\theta^2
    +\text{sin}^2\theta(d\phi-\omega\, dt)^2\right],
\end{equation}
where $N,\mu,K,\text{and}\,\, \omega$ are all functions of $r$
and $\theta$; $\omega$ is the angular velocity $d\phi/dt$.  More
precisely,
\[
    \omega=\frac{d\phi}{dt}=\frac{d\phi/d\tau}{dt/d\tau}=
    \frac{u^\phi}{u^t},
\]
referred to in Ref.~\cite{MTW73} as the ``angular velocity relative
to the asymptotic rest frame."  The connection between the metric 
due to a bounded rotating source and the mass and angular momentum
of the source is discussed in Ref.~\cite{jI85}.

To make our solution as general as possible, we will assume that
$\omega=\omega(r,\theta,t)$ is time-dependent, so that the wormhole 
can no longer be called stationary.  The reason for proposing this
model is a practical one: if an advanced civilization were to
succeed in constructing such a wormhole, it is likely to be 
aspherical and the rate of rotation likely to be varied.  Accordingly,
we will write our line element as follows:
\begin{multline}\label{E:line4}
     ds^2=-e^{2\lambda(r,\theta)}dt^2\\+e^{2\mu(r,\theta)}dr^2
     +\left[K(r,\theta)\right]^2r^2\left[d\theta^2-\text{sin}^2
     \theta(d\phi-\omega(r,\theta,t)dt)^2\right].
\end{multline}
Here $K(r,\theta)$  is a positive dimensionless function of $r$ 
such that $Kr$ determines the proper radial distance at $(r,\theta)$ in
the usual manner.  In other words, $2\pi(Kr)\text{sin}\,\theta$ is the
proper circumference of the circle through $(r,\theta)$.

So far nothing has been said about the shape function.  Recall that
the wormhole geometry may be conveniently described by means of an 
embedding diagram in three-dimensional Euclidean space at a fixed
moment in time and for a fixed value of $\theta$, the equatorial slice
$\theta=\pi/2$~\cite{MT88,aD01}.  The resulting surface of revolution
has the parametric form
\[
    f(r,\phi)=(r\,\text{cos}\,\phi,r\,\text{sin}\,\phi,
    z(r,\theta_1)),
\]
where $z=z(r,\theta)$ is some function of $r$ and $\theta$ and $\theta$
is momentarily held fixed at $\theta_1$.  As usual, we think of the 
surface as connecting two asymptotically flat universes.  The radial
coordinate decreases from $+\infty$ in the ``upper" universe to a
minimum value $r=r_0$ at the throat, and then increases again to
$+\infty$ in the ``lower" universe.

Since the embedding surface must have a vertical tangent at the throat
for any value of $\theta$, we require that 
\[
     \lim_{r \to r_0+}\frac{dz}{dr}=+\infty,
\]
while $\lim_{r \to \infty}dz/dr=0$, the meaning of asymptotic flatness.
Returning to the line element (\ref{E:line4}), we further assume that 
for any fixed $\theta$, $\mu(r,\theta)$ has a vertical asymptote at
$r=r_0$: $\lim_{r \to r_0+}\mu(r,\theta)=+\infty$.  Also, 
$\mu(r,\theta)$ is a twice differentiable function of $r$ and $\theta$ 
(as is $\lambda(r,\theta)$) and is a strictly decreasing function of
$r$ with $\lim_{r \to \infty}\mu(r,\theta)=0$.

These requirements are met by $z=z(r,\theta)$ for any fixed $\theta$
such that
\[
      \frac{dz}{dr}=\sqrt{e^{2\mu(r,\theta)}-1}
\] 
(for the upper universe).  Furthermore, $d^2z/dr^2<0$ near the throat
(since $d\mu(r,\theta_1)/dr<0$), as required by the ``flaring out"
condition in Ref.~\cite{MT88}.  The shape function is now defined by
\[
      e^{2\mu(r,\theta)}=\frac{1}{1-\frac{b(r,\theta)}{r}}.
\]
It follows that
\[
       b(r,\theta)=r\left(1-e^{-2\mu(r,\theta)}\right).
\]
Finally, at the throat itself, $b$ must be independent of $\theta$.
It is readily checked that $\partial b/\partial \theta=0$ at
$r=r_0$.

\section{The solution}\label{S:solution}
Let us write the line element (\ref{E:line4}) in slightly more compact form:
\begin{equation}\label{E:line5}
    ds^2=-e^{2\lambda}dt^2+e^{2\mu}dr^2+K^2r^2\left[d\theta^2
    +\text{sin}^2\theta(d\phi-\omega\,dt)^2\right].
\end{equation}

To make the analysis tractable, we choose an orthonormal basis 
$\{e_{\hat{\alpha}}\}$ which is dual to the following 1-form basis:
\begin{equation}\label{E:oneform1}
    \theta^0=e^{\lambda}\, dt,\qquad \theta^1=e^{\mu}\,dr, 
     \qquad\theta^2=Kr\,d\theta
\end{equation}
and
\begin{equation}\label{E:oneform2}
      \theta^3=Kr\,
\,\text{sin}\,\theta(d\phi-\omega\,dt).
\end{equation}
(See also Ref.~\cite{vK98}.)  As a result,
\begin{equation}\label{E:oneform3}
     dt=e^{-\lambda}\,\theta^0,\qquad dr=e^{-\mu}\,\theta^1,
     \qquad d\theta=\frac{1}{Kr}\theta^2,
\end{equation}
and
\begin{equation}\label{E:oneform4}
    d\phi=\frac{1}{Kr\,\text{sin}\,\theta}\theta^3+\omega
    \,e^{-\lambda}\,\theta^0.
\end{equation}
Furthermore, 
\[
    ds^2=-(d\theta^0)^2+(d\theta^1)^2+(d\theta^2)^2+(d\theta^3)^2.
\]

Since the orthonormal basis is itself rotating, some information 
regarding $\omega$ is going to be lost.  We will briefly return to 
this topic at the end of the last section.  (In particular, it
will be shown that the magnitude of $\omega$ must be restricted.)
To do so, we need the components of the fundamental metric tensor
$\{g_{\alpha \beta}\}$, as well as $\{g^{\alpha \beta}\}$, in the 
$(t,r,\theta,\phi)$-coordinate system:
\begin{equation}\label{E:metrictensor}
    g_{tt}=-e^{2\lambda}+K^2r^2\omega^2\,\text{sin}^2\theta,\qquad
    g_{t\phi}=-K^2r^2\omega\,\text{sin}^2\theta,
\end{equation}

\begin{equation*}
    g_{rr}=e^{2\mu},\qquad g_{\theta\theta}=K^2r^2,\qquad 
       g_{\phi\phi}=K^2r^2\,\text{sin}^2\theta,
\end{equation*}
and
\begin{equation}\label{E:inverse}
    g^{tt}=-e^{-2\lambda},\qquad g^{t\phi}=-\omega\, e^{-2\lambda},\qquad
      g^{rr}=e^{-2\mu},
\end{equation}
\begin{equation*}
    g^{\theta\theta}=\frac{1}{K^2r^2},\qquad g^{\phi\phi}=
    \frac{1}{K^2r^2\,\text{sin}^2\theta}-\omega^2e^{-2\lambda}.
\end{equation*}
The last component, $g^{\phi\phi}$, bears a striking similarity
to $d\phi$ in Eq.~(\ref{E:oneform4}).

To obtain the curvature 2-forms and the components of the
Riemann curvature tensor, we use the method of differential
forms (Ref.~\cite{HT90}).  To that end we calculate the following
exterior derivatives in terms of $\theta^i$:
\[
   d\theta^0=\frac{\partial\lambda}{\partial r}e^{-\mu}
   \,\theta^1\wedge\theta^0+\frac{1}{Kr}\frac{\partial\lambda}
      {\partial\theta}\,\theta^2\wedge\theta^0,\qquad d\theta^1=
       \frac{1}{Kr}\frac{\partial\mu}{\partial\theta}\,
          \theta^2\wedge\theta^1,
\]
\[
   d\theta^2=\left(\frac{1}{r}e^{-\mu}+\frac{1}{K}\frac
     {\partial K}{\partial r}e^{-\mu}\right)
        \,\theta^1\wedge\theta^2,
\]
and
\begin{multline*}
   d\theta^3=-Kr\frac{\partial\omega}{\partial r}e^{-\lambda}
   e^{-\mu}\text{sin}\,\theta\,\,\,\theta^1\wedge\theta^0
   -\frac{\partial\omega}{\partial\theta}e^{-\lambda}
   \text{sin}\,\theta\,\,\,\theta^2\wedge\theta^0\\
     +\left(\frac{1}{r}e^{-\mu}+\frac{1}{K}\frac{\partial K}
      {\partial r}e^{-\mu}\right)\,\theta^1\wedge\theta^3
      +\left(\frac{1}{Kr}\text{cot}\,\theta+\frac{1}{K^2r}
      \frac{\partial K}{\partial\theta}\right)\,\theta^2
       \wedge\theta^3.
\end{multline*}

The connection 1-forms $\omega^i_{\phantom{i}\,\,k}$ have the
symmetry
\[
    \omega^0_{\phantom{i}\,\,i}=\omega^i_{\phantom{0}0}
    \;(i=1,2,3),\;\text{and}\;\omega^i_{\phantom{j}j}=
     -\omega^j_{\phantom{i}\,i}\;(i,j=1,2,3, i\ne j)
\]
and are related to the basis $\theta^i$ by
\[
   d\theta^i=-\omega^i_{\phantom{k}k}\wedge\theta^k.
\]
The solution of this system is found to be
\begin{align*}
   \omega^0_{\phantom{0}1}&=\frac{1}{2}Kr\frac{\partial\omega}
   {\partial r}e^{-\lambda}e^{-\mu}\text{sin}\,\theta\;
   \theta^3+\frac{\partial\lambda}{\partial r}e^{-\mu}\;\theta^0,\\  
   \omega^0_{\phantom{0}2}&=\frac{1}{Kr}\frac{\partial\lambda}
   {\partial\theta}\;\theta^0+\frac{1}{2}\frac{\partial\omega}
   {\partial\theta}e^{-\lambda}\text{sin}\,\theta\;\theta^3,\\
   \omega^0_{\phantom{0}3}&=\frac{1}{2}Kr\frac{\partial\omega}
    {\partial r}e^{-\lambda}e^{-\mu}\text{sin}\,\theta\;\theta^1
      +\frac{1}{2}\frac{\partial\omega}{\partial\theta}
       e^{-\lambda}\text{sin}\,\theta\;\theta^2,\\
    \omega^1_{\phantom{0}2}&=\frac{1}{Kr}\frac{\partial\mu}
    {\partial\theta}\;\theta^1-\left(\frac{1}{r}e^{-\mu}
     +\frac{1}{K}\frac{\partial K}{\partial r}
        e^{-\mu}\right)\;\theta^2,\\
   \omega^1_{\phantom{0}3}&=\frac{1}{2}Kr\frac{\partial\omega}
    {\partial r}e^{-\lambda}e^{-\mu}\text{sin}\,\theta\;\theta^0
      -\left(\frac{1}{r}e^{-\mu}+\frac{1}{K}\frac{\partial K}
       {\partial r}e^{-\mu}\right)\;\theta^3,\\
     \omega^2_{\phantom{0}3}&=\frac{1}{2}\frac{\partial\omega} 
     {\partial\theta}e^{-\lambda}\text{sin}\,\theta\;\theta^0
        -\left(\frac{1}{Kr}\text{cot}\,\theta+\frac{1}{K^2r}
         \frac{\partial K}{\partial\theta}\right)\;\theta^3.
\end{align*}

The curvature 2-forms $\Omega^i_{\phantom{j}j}$ are calculated
directly from the Cartan structural equations
\[
    \Omega^i_{\phantom{j}j}=d\omega^i_{\phantom{j}j} +\omega^i
     _{\phantom{j}k}\wedge\omega^k_{\phantom{j}j}.
\]
The results are given in the Appendix.  Since the components 
of the Riemann curvature tensor can be read off directly using 
the formula
\[
   \Omega^i_{\phantom{j}j}=-\frac{1}{2}R_{mnj}^{\phantom{mnj}i}
    \;\theta^m\wedge\theta^n,
\]
there is no need to list them explicitly.  As an example, suppose
we let $m=0$ and $n=1$ in the equation
\[
    \Omega^0_{\phantom{0}1}=-\frac{1}{2}R_{mn1}^{\phantom{mnn}0}
      \;\theta^m\wedge\theta^n.
\]
Then
\begin{multline*}
   \Omega^0_{\phantom{0}1}=-\frac{1}{2}R_{011}^{\phantom{000}0}
    \;\theta^0\wedge\theta^1-\frac{1}{2}R_{101}^{\phantom{000}0}
    \;\theta^1\wedge\theta^0\\=-R_{011}^{\phantom{000}0}\;
     \theta^0\wedge\theta^1=R_{011}^{\phantom{000}0}\;
         \theta^1\wedge\theta^0.
\end{multline*}
Thus $R_{011}^{\phantom{000}0}=A(1,0)$ in the Appendix.  (As in
Ref.~\cite{HT90}, we omit the hats whenever numerical indices are
used.)

\section{WEC violation}\label{S:WECviolation}
As with any traversable wormhole, we would expect a violation of the 
weak energy condition near the throat: $T_{\hat{\alpha}\hat{\beta}}
\mu^{\hat{\alpha}}\mu^{\hat{\beta}}\ge0$ for all null vectors.  As in
Morris and Thorne~\cite{MT88} and Roman~\cite{tR93} we use a radial 
outgoing null vector $\mu^{\hat{\alpha}}=(\mu^{\hat{t}},\mu^{\hat{r}}
,0,0)=(1,1,0,0)$.  In our orthonormal frame we expect to have the 
usual stress-energy components $T_{\hat{t}\hat{t}}$, 
$T_{\hat{r}\hat{r}}$, $T_{\hat{\theta}\hat{\theta}}$, $T_{\hat{\phi}
\hat{\phi}}$, as well as $T_{\hat{t}\hat{\phi}}$, which represents
the rotation of the matter distribution \cite{eT98}.

From $G_{\hat{\alpha}\hat{\beta}}=R_{\hat{\alpha}\hat{\beta}}-\frac{1}
{2}Rg_{\hat{\alpha}\hat{\beta}}$ and $R_{ab}=R_{acb}^{\phantom{abc}c}$,
we have $G_{00}+G_{11}=R_{00}+R_{11}=R_{011}^{\phantom{000}0}+
R_{022}^{\phantom{000}0}+R_{033}^{\phantom{000}0}-R_{011}^{\phantom
{000}0}-R_{122}^{\phantom{000}1}-R_{133}^{\phantom{000}1}$.  A short
calculation yields
\begin{multline}\label{E:WEC}
   8\pi\left(T_{\hat{t}\hat{t}}+T_{\hat{r}\hat{r}}\right)=R_{\hat{t}
    \hat{t}}+R_{\hat{r}\hat{r}}\\=\frac{2}{r}e^{-2\mu}\left(
    \frac{\partial\lambda}{\partial r}+\frac{\partial\mu}{\partial r}
    \right) +\frac{2}{K}\frac{\partial K}{\partial r}
     e^{-2\mu}\left(\frac{\partial\lambda}{\partial r}+
     \frac{\partial\mu}{\partial r}\right)\\
           +\frac{1}{K^2r^2}\left(\frac{\partial\lambda}{\partial\theta}
          -\frac{\partial\mu}{\partial\theta}\right)\text{cot}\,\theta
     +\frac{1}{K^2r^2}\left(\frac{\partial^2\lambda}{\partial\theta^2}
      -\frac{\partial^2\mu}{\partial\theta^2}\right)\\
             +\frac{1}{K^2r^2}\left[\left(\frac{\partial\lambda}
              {\partial\theta}\right)^2-\left(\frac{\partial\mu}
              {\partial\theta}\right)^2\right]
      +\left(-\frac{4}{Kr}\frac{\partial K}{\partial r}e^{-2\mu}\right)\\
          +\left(-\frac{2}{K}\frac{\partial^2K}{\partial r^2}e^{-2\mu}\right)
       +\left[-\frac{1}{2}\left(\frac{\partial\omega}
          {\partial\theta}\right)^2
          e^{-2\lambda}\text{sin}^2\theta\right].
\end{multline}
(We have used the Einstein field equations $G_{\hat{\alpha}\hat{\beta}}
=8\pi T_{\hat{\alpha}\hat{\beta}}$.)

To put this rather long expression in perspective, consider the static
spherical case \cite{pK99}, where only the first term survives:
\begin{equation}\label{E:WECstatic}
    T_{\hat{t}\hat{t}}+T_{\hat{r}\hat{r}}=\rho-\tau=\frac{1}{8\pi}
    \left(\frac{2}{r}\right)e^{-2\mu}\left(\frac{d\lambda}{dr}+
     \frac{d\mu}{dr}\right).
\end{equation}
Referring to Sect.~\ref{S:metric}, recall that the throat corresponds to
the value $r=r_0$.  So 
if $\mu$ is a smooth function of $r$ 
and given that $\lim_{r \to r_0+}\mu(r)=+\infty$, it follows that 
$\lim_{r \to r_0+}d\mu/dr=-\infty$. In Ref.~\cite{pK99}, 
$\lambda(r)=-\kappa/r$, $\kappa>0$, so that $d\lambda/dr=\kappa/r^2$, making $\rho
-\tau$ negative near the throat.  That a violation of the WEC cannot be 
avoided regardless of the choice of $\lambda(r)$ can be seen 
geometrically.  For if $\lim_{r \to r_0+}d\lambda/dr$ is positive and
finite, then the sum on the right side of Eq.~(\ref{E:WECstatic}) is
negative near the throat.  This might be avoided if 
$\lim_{r \to r_0+}d\lambda/dr=+\infty$.  But then $\lim_{r \to r_0+}
\lambda(r)=-\infty$ and $e^{2\lambda(r)}\rightarrow0$, which yields
an event horizon.

The main goal in this section is to show that for a rotating axially
symmetric wormhole the WEC violation is, in principle, much less
severe than for the static spherically symmetric case by a suitable
choice of $\lambda$ and $\mu$.  The functions were chosen primarily
for convenience, keeping the analysis simple and accommodating the next
section at the same time.

Taking $k, A, \text{and} \,B$ to be constants, let 
\begin{equation}\label{E:redshift}
    \lambda(r,\theta)= -\frac{k}{r}\left[-\frac{1}{3}\left(\frac{\pi}{2}
    -\theta\right)^3+A\right],\quad 0<\theta\le\frac{\pi}{2},\quad k>0,
\end{equation}
where $A$ is large enough to keep the expression inside the
brackets positive, and
\begin{equation}\label{E:shape}
    \mu(r,\theta)=\frac{k\epsilon}{r-r_0}\left[\left(\frac{\pi}{2}-\theta
    \right)+B\right],\quad 0<\theta\le\frac{\pi}{2},\quad k,B>0,
\end{equation}
where $\epsilon>0$ is a small constant. On the interval $[\pi/2,\pi)$ 
each function is
defined to be the ``mirror image,"i.e., $\lambda$ and $\mu$ are symmetric
about $\theta=\pi/2$.  So the discussion may be confined to the
interval $(0,\pi/2]$.  

The partial derivatives are listed next for easy reference:
\begin{align*}
   \frac{\partial\lambda}{\partial r}&=
      \frac{k}{r^2}\left[-\frac{1}{3}\left(\frac{\pi}{2}
          -\theta\right)^3+A\right],
      &\frac{\partial\mu}{\partial r}&=-\frac{k\epsilon}
             {(r-r_0)^2}\left [\left(\frac{\pi}{2}-\theta\right)+B\right ],\\
       \frac{\partial\lambda}{\partial\theta}&=
           -\frac{k}{r}\left(\frac{\pi}{2}-\theta\right)^2,
        &\frac{\partial\mu}{\partial\theta}&=-\frac{k\epsilon}
          {r-r_0},\\
        \frac{\partial^2\lambda}{\partial\theta^2}&=
            \frac{2k}{r}\left(\frac{\pi}{2}-\theta\right),
        &\frac{\partial^2\mu}{\partial\theta^2}&=0.
\end{align*}

On the right side of Eq.~(\ref{E:WEC}), the first two terms are similar
to those in Eq.~(\ref{E:WECstatic}).  The first term is therefore 
negative.  The fourth term is stricly positive for $\theta\ne\pi/2$, 
while the third term 
is positive near the throat.  In the fifth term the expression
\[
   \left(\frac{\partial\lambda}{\partial\theta}\right)^2
  -\left(\frac{\partial\mu}{\partial\theta}\right)^2
     =\frac{k^2}{r^2}\left(\frac{\pi}{2}-\theta\right)^4
         -\frac{k^2\epsilon^2}{(r-r_0)^2}
\]
is similar to the first term but the quadratic factor $\epsilon^2$ 
reduces the size of the second term, thereby making the
fifth term less harmful.

To study the effect of $K(r,\theta)$, we adopt a function similar to
one suggested by Teo \cite{eT98},
\[
    K(r,\theta)=1+\frac{(4a\,\,\text{sin}\,\theta)^2}{r}.
\]
Now the second and sixth terms are positive, as well.  The seventh term
is close to zero near the throat, thanks to the factor $e^{-2\mu}$, and
so is completely overshadowed by the third and fourth terms.

The last term is more of a problem, being strictly negative.  However,
for Teo's choice, $\omega=2a/r^3$, where $a$ is the total angular
momentum of the wormhole, the last term is zero.  Requiring $\omega$
to be independent of $\theta$ may be unavoidable, a conclusion also
reached by Khatsymovsky~\cite{vK98}, who states that for a macroscopic 
wormhole to exist, the angular velocity must be independent of $\theta$.

With the last term eliminated, we see that a drastic reduction in the
energy condition violation is indeed possible, at least in principle.

\emph{Remark:}  We will see in the next section that the absolute value
of the first term exceeds that of the second term.  Otherwise the WEC
violation would appear to have been eliminated completely.

\section{Traversibility conditions}\label{S:traversability}
Another area in which improvements over the static spherically
symmetric case are possible is in the study of tidal constraints.  In
particular, for an infalling radial observer the components of the
Riemann curvature tensor are found relative to the following
orthonormal basis (from the usual Lorentz transformations):

\begin{equation}\label{E:Lorentz}
    e_{\hat{0}'}=\gamma e_{\hat{t}}\mp\gamma\left(\frac{v}{c}\right)
       e_{\hat{r}},\quad e_{\hat{1}'}=\mp\gamma e_{\hat{r}}
        +\gamma\left(\frac{v}{c}\right)e_{\hat{t}},\quad
    e_{\hat{2}'}=e_{\hat{\theta}},\quad 
          e_{\hat{3}'}=e_{\hat{\phi}}.
\end{equation}
A traveler should not experience any tidal forces larger than those
on Earth.  As outlined in Ref.~\cite{MT88}, the radial tidal constraint
is given by
\begin{equation*}
   \left|R_{\hat{1}'\hat{0}'\hat{1}'\hat{0}'}\right|\leq
     \frac{g_\oplus}{c^2\times 2\,\text{m}}\approx
       \frac{1}{(10^8\, \text{m})^2},
\end{equation*}
assuming an observer $2\,\text{m}$ tall.  We have
\begin{multline}\label{E:radial}
    \left|R_{\hat{1}'\hat{0}'\hat{1}'\hat{0}'}\right|
         =\left|R_{\hat{r}\hat{t}\hat{r}\hat{t}}\right|
    =\left|e^{-2\mu}\left[\frac{\partial^2\lambda}
       {\partial r^2}-\frac{\partial\lambda}{\partial r}
       \frac{\partial\mu}{\partial r}+
           \left(\frac{\partial\lambda}{\partial r}\right)^2\right]
              \right.\\
    \left.+\left(-\frac{3}{4}\right)K^2r^2\left(\frac{\partial\omega}{\partial r}
      \right)^2e^{-2\lambda}e^{-2\mu}\text{sin}^2\theta
        +\frac{1}{K^2r^2}\frac{\partial\lambda}
           {\partial\theta}\frac{\partial\mu}{\partial\theta}\right|.
\end{multline}

In the static spherically symmetric case only the first term survives.
The second term, which involves the angular velocity, reduces the size
of $\left|R_{\hat{r}\hat{t}\hat{r}\hat{t}}\right|$: the first term, 
\[
     e^{-2\mu}\left[\frac{\partial^2\lambda}{\partial r^2}+
       \left(-\frac{\partial\lambda}{\partial r}\frac{\partial\mu}
         {\partial r}\right)
          +\left(\frac{\partial\lambda}{\partial r}\right)^2\right],
\]
is positive near the throat because the positive middle term inside the
brackets contains the factor $(r-r_0)^2$ in the denominator.  For the 
same reason the first term is larger than the absolute value of the
second (near the throat).  Since the second term is negative, the net 
result, so far, is a reduction in the size of 
$\left|R_{\hat{r}\hat{t}\hat{r}\hat{t}}\right|$.  

Concerning the last term, we need to remember that our wormhole is not
spherically symmetric.  The tidal forces experienced may therefore 
depend on the direction of approach.  So we must ask the traveler to 
approach the throat in the equatorial plane $\theta=\pi/2$, and here 
$\partial\lambda/\partial\theta=0$.

To study the first of the lateral tidal constraints, 
$\left|R_{\hat{2}'\hat{0}'\hat{2}'\hat{0}'}\right|
      \leq(10^8\,\text{m})^{-2}$, we have from Eq.~(\ref{E:Lorentz}),

\begin{equation*}
   \left|R_{\hat{2}'\hat{0}'\hat{2}'\hat{0}'}\right|
    =\gamma^2\left|R_{\hat{\theta}\hat{t}\hat{\theta}\hat{t}}\right |
     +\gamma^2\left(\frac{v}{c}\right)^2\left|
          R_{\hat\theta\hat{r}\hat{\theta}\hat{r}}\right|.
\end{equation*}
As usual, this is a constraint on the velocity of the traveler.  It is
assumed in Ref.~\cite{MT88} that the spaceship decelerates until it
comes to rest at the throat.  So only the first term needs to be 
examined:
\begin{multline}\label{E:lateral1}
   \left|R_{\hat{\theta}\hat{t}\hat{\theta}\hat{t}}\right|
       =\left|R_{022}^{\phantom{000}0}\right|\\
   =\left|\frac{1}{r}\frac{\partial\lambda}{\partial r}e^{-2\mu}
      +\frac{1}{K^2r^2}\frac{\partial^2\lambda}{\partial\theta ^2}
       +\frac{1}{K^2r^2}\left(\frac{\partial\lambda}{\partial
            \theta}\right)^2
    -\frac{1}{K^3r^2}\frac{\partial K}{\partial\theta}
         \frac{\partial\lambda}{\partial\theta}\right.\\
     \left.+\frac{1}{K}\frac{\partial K}{\partial r}
        \frac{\partial\lambda}{\partial r}e^{-2\mu}
     -\frac{3}{4}\left(\frac{\partial\omega}
        {\partial\theta}\right)^2e^{-2\lambda}\text{sin}^2\theta\right|.
\end{multline}

In the static spherically symmetric case only the positive first term
survives.  Once again requiring that the traveler approach the throat
in the equatorial plane $\theta=\pi/2$, the next three terms are zero.
 The last term is also zero due to the earlier requirement 
$\partial\omega/\partial\theta=0$.
The fourth term is negative, and for our choice of $K(r,\theta)$
\begin{equation*}
   \left|\frac{1}{K}\frac{\partial K}{\partial r}
       \frac{\partial\lambda}{\partial r}e^{-2\mu}\right|
   <\left|\frac{1}{r}\frac{\partial\lambda}
      {\partial r}e^{-2\mu}\right|.
\end{equation*}
We therefore have a reduction in the size of
$\left|R_{\hat{\theta}\hat{t}\hat{\theta}\hat{t}}\right|$.

For the remaining lateral tidal constraint we need to examine
\begin{multline}\label{E:lateral2}
    \left|R_{\hat{\phi}\hat{t}\hat{\phi}\hat{t}}\right|
    =\left|R_{033}^{\phantom{000}0}\right |
    =\left|\frac{1}{r}\frac{\partial\lambda}{\partial r}e^{-2\mu}
        +\frac{1}{K}\frac{\partial K}{\partial r} 
          \frac{\partial\lambda}{\partial r}
           e^{-2\mu}\right.\\
    +\frac{1}{4}K^2r^2\left(\frac{\partial\omega}{\partial r}\right)^2
        e^{-2\lambda}e^{-2\mu}\text{sin}^2\theta
    +\frac{1}{K^2r^2}\frac{\partial\lambda}{\partial\theta}
        \text{cot}\,\theta\\
     \left.+\frac{1}{K^3r^2}\frac{\partial K}{\partial\theta}
        \frac{\partial\lambda}{\partial\theta}
         +\frac{1}{4}\left(\frac{\partial\omega}
            {\partial\theta}\right)^2e^{-2\lambda}\text{sin}^2\theta
         \right|.    
\end{multline}
The first two terms appeared in the other lateral constraint; the
result is a reduction in the size of 
$\left|R_{\hat{\phi}\hat{t}\hat{\phi}\hat{t}}\right|$.  Unfortunately, 
the next term is positive.  Although small due to the factor 
$e^{-2\mu}$, the overall result is hard to quantify and may actually
be less favorable than the corresponding static case.  (It helps that
the next two terms are zero for $\theta=\pi/2$, while the last term
vanishes due to the requirement $\partial\omega/\partial\theta=0$.)

\section{The effect of rotation}\label{S:rotation}
In proposing a model with a time-dependent $\omega$, i.e., an angular
acceleration or deceleration, it was hoped that the findings in the
last two sections would be strengthened.  As it turns out, however, 
none of the derivatives with respect to time occurring in the
curvature 2-forms found their way into the earlier calculations.

This failure suggests that the effect of rotation be studied from a
different perspective.  Since we are using a rotating basis, some
information regarding $\omega$ was lost: no $\omega$'s appear in any
of the curvature 2-forms, although the derivatives do.  So it may
be useful to examine the weak energy condition violation relative to
the $(t,r,\theta,\phi)$-coordinate system.

The expression for $T_{tt}+T_{rr}$, calculated by the traditional method
using the fundamental metric tensor (Equations (\ref{E:metrictensor}) and
(\ref{E:inverse})), contain the old terms in Eq. (\ref{E:WEC}), but not, 
of course, in the orthonormal frame.  The new terms all contain $\omega$:
\[
   8\pi(T_{tt}+T_{rr})=\text{old terms}+R_{\text{new}},
\]
where
\begin{multline*}
   R_{\text{new}}=\\-\frac{1}{2}\omega K^3r^4\frac{\partial K}{\partial r}
       \frac{\partial\omega}{\partial r}e^{-2\lambda}e^{-2\mu}
           \text{sin}^2\theta
      +\left(-\frac{1}{2}\right)\omega K^4r^3\frac{\partial\omega}{\partial r}
           e^{-2\lambda}e^{-2\mu}\text{sin}^2\theta\\
      -\omega Kr^2\frac{\partial K}{\partial\theta}\frac{\partial\omega}
          {\partial\theta}e^{-2\lambda}\text{sin}^2\theta
       -\frac{1}{2}\omega K^2r^2\frac{\partial^2\omega}
           {\partial\theta^2}e^{-2\lambda}\text{sin}^2\theta\\
       +\frac{1}{2}\omega K^2r^2\frac{\partial\omega}{\partial\theta}
            \frac{\partial\lambda}{\partial\theta}e^{-2\lambda} 
             \text{sin}^2\theta
       -\frac{3}{2}\omega K^2r^2\frac{\partial\omega}{\partial\theta}
             e^{-2\lambda}\text{sin}\,\theta\,\text{cos}\,\theta.
\end{multline*}
$R_{\text{new}}$ obviously vanishes if $\omega$ does.  The last four 
terms vanish if we assume, as before, that $\partial\omega/\partial
\theta=0$.  With the functions used earlier, the first term is negative
and the second positive.  Unless $\partial K/\partial r$ is very small,
this result is, once again, hard to quantify.

The situation is rather different if we return to the assumption of
spherical symmetry, while retaining $\omega$, assumed to be positive.
Then $K=1$ and $\lambda$ and $\mu$ are independent of $\theta$.  In 
the $(t,r,\theta,\phi)$-coordinate system, using the outgoing null
vector $(1,1,0,0)$, we have
\begin{multline}\label{E:WECreduced}
   \rho-\tau=re^{-2\mu}\left(\frac{\partial\lambda}{\partial r}
       +\frac{\partial\lambda}{\partial r}\text{sin}^2\theta\right)
   +re^{-2\mu}\left(\frac{\partial\mu}{\partial r}
        +\frac{\partial\mu}{\partial r}\text{sin}^2\theta\right)\\
   +\left(-\frac{1}{2}\right)\omega r^3\frac{\partial\omega}{\partial r}
        e^{-2\lambda}e^{-2\mu}\text{sin}^2\theta.
\end{multline}
Since $\omega>0$, the last term is positive.  So if $\omega$ is large,
the WEC violation is much reduced.

While the resulting reduction in the WEC violation is a welcome 
surprise, some words of caution are in order.  The last term in Eq.
(\ref{E:WECreduced}) suggests that if $\omega$ is large enough, the
WEC violation can be eliminated altogether.  But as explained by
Teo \cite{eT98}, for a rapidly rotating wormhole it may not be
possible to use a radially outgoing null vector since the $g_{tt}$
component of the fundamental metric tensor may no longer be negative,
as can be seen from the line element, Eq. (\ref{E:line5}).

\section{Additional considerations}\label{S:additional}
It was pointed out in a recent paper by Perez Bergliaffa and 
Hibberd \cite{PB02} that the metric (\ref{E:line4})
used in this paper may require further restrictions.  In particular,
it is shown that a wormhole of the type studied by Teo~\cite{eT98}
cannot be generated by a perfect fluid or by a fluid with anisotropic
stresses.  In the first case the condition
\begin{equation}\label{E:perfect}
   G_{12}=0
\end{equation}
is violated and in the second case, 
\begin{equation}\label{E:anisotropic}
   G_{00}+G_{33}\ge 2\,G_{03}.
\end{equation}

These conditions seem to be met if the metric (\ref{E:line4}) includes
the functions $\lambda$ and $\mu$ used in this paper, that is,
Equations (\ref{E:redshift}) and (\ref{E:shape}), respectively.

For example, to check Condition (\ref{E:perfect}), a short calculation
yields
\begin{multline*}
   G_{12}= \frac{1}{2}Kr\frac{\partial\omega}{\partial r}
    \frac{\partial\omega}{\partial\theta}e^{-2\lambda}e^{-\mu}
    \text{sin}^2\theta
      +\frac{1}{Kr^2}\left(\frac{\partial\lambda}{\partial\theta}
      +\frac{\partial\mu}{\partial\theta}\right)e^{-\mu}\\
    +\frac{1}{K^2r}\frac{\partial K}{\partial r}
        \left(\frac{\partial\lambda}{\partial\theta}
         +\frac{\partial\mu}{\partial\theta}\right)e^{-\mu}
     -\frac{1}{Kr}\frac{\partial\lambda}{\partial r}
      \left(\frac{\partial\lambda}{\partial\theta}-\frac{\partial\mu}
         {\partial\theta}\right)e^{-\mu}\\-\frac{1}{Kr}
            \frac{\partial^2\lambda}{\partial r\partial\theta}e^{-\mu}
    -\frac{1}{K^2r}\frac{\partial^2K}{\partial r\partial\theta}e^{-\mu}
          +\frac{1}{K^3r}\frac{\partial K}{\partial r}
          \frac{\partial K}{\partial\theta}e^{-\mu}.
\end{multline*}
Each term contains $e^{-\mu}$, so that $G_{12}\approx 0$ near the throat.

Even more favorable is the outcome of the check on Condition 
(\ref{E:anisotropic}): five of the terms in $G_{03}$ contain the factor
$\partial\omega/\partial\theta$, which is equal to zero.  The remaining
terms all contain the factor $e^{-2\mu}$, so that $G_{03}\approx 0$
near the throat.  The left side, $G_{00}+G_{33}$, contains only two
terms that are neither strictly positive, nor zero, nor contain the factor 
$e^{-2\mu}$.  But these two terms are strongly overshadowed by several
terms that \emph{are} strictly positive.  So Condition 
(\ref{E:anisotropic}) is easily met in the vicinity of the throat.

Since the vicinity of the throat is the only region that really
matters, we can construct (in the usual manner)
a solution with a suitable radial cut-off of the stress-energy tensor.
Our solution can then be joined, at least in principle,
to an external solution also satisfying the desired conditions.  (For a
discussion of the required junction conditions, see Ref.~\cite{MTW73}.)

It is indeed surprising that $\lambda$ and $\mu$, which were chosen for
entirely different reasons, are sufficient for satisfying Conditions
(\ref{E:perfect}) and (\ref{E:anisotropic}), thereby overcoming the 
objections raised to the wormhole in Teo~\cite{eT98}.  Unfortunately,
Ref~\cite{PB02} discusses other conditions, some of which are not so 
easily checked. So it may still be necessary to ``incorporate more
realistic (in the astrophysical sense) features, for example heat flux"
(Ref~\cite{PB02}).  Future investigations of these issues could be
extended to include stability conditions, especially for rotating
wormholes, since such conditions may also require ``more realistic
features."

\section{Conclusion}
In this paper the MT wormhole solution was generalized to wormholes 
which are both rotating and axially symmetric, i.e., symmetric with 
respect to the axis of rotation.  It was concluded that the
unavoidable violation of the weak energy condition is less severe
than in the spherically symmetric case.  The radial tidal constraint
is more easily met due to the rotation.  An improvement was also 
found in one of the lateral tidal constraints.  Making the angular
velocity $\omega$ time-dependent does not help since none of the
time-derivatives in the Appendix appeared in the calculations.  
Furthermore, $\omega$ must be independent of $\theta$, in agreement
with Ref.~\cite{vK98}.  Finally, 
the magnitude of the angular velocity may have little effect on the
WEC violation for an axially symmetric wormhole.  In the spherically
symmetric case, however, a rapid rotation will result in a reduction 
in the WEC violation, as long as $\omega$ is not excessively large.

\section*{Appendix}
This appendix lists the curvature 2-forms $\Omega^i_{\phantom{j}j}$.
\begin{multline*}
   \Omega^0_{\phantom{0}1}=A(1,0)\,\theta^1\wedge\theta^0
       +A(2,0)\,\theta^2\wedge\theta^0+A(3,0)\,\theta^3\wedge
       \theta^0+A(3,1)\,\theta^3\wedge\theta^1\\
    +A(3,2)\,\theta^3\wedge\theta^2,
\end{multline*}
where
\begin{multline*}
   A(1,0)=e^{-2\mu}\left[\frac{\partial^2\lambda}{\partial r^2}
      -\frac{\partial\lambda}{\partial r}\frac{\partial\mu}
       {\partial r} +\left(\frac{\partial\lambda}{\partial r}
          \right)^2\right]\\
    -\frac{3}{4}K^2r^2\left(\frac{\partial\omega}{\partial r}
       \right)^2e^{-2\lambda}e^{-2\mu}\text{sin}^2\theta
    +\frac{1}{K^2r^2}\frac{\partial\lambda}{\partial\theta}
        \frac{\partial\mu}{\partial\theta},
\end{multline*}
\begin{multline*}
   A(2,0)=\frac{1}{Kr}e^{-\mu}\left(\frac{\partial^2\lambda}
     {\partial r\partial\theta}-\frac{\partial\lambda}
         {\partial r}\frac{\partial\mu}{\partial\theta}
         +\frac{\partial\lambda}{\partial r}\frac{\partial\lambda}
       {\partial\theta}\right)\\
    -\frac{3}{4}Kr\frac{\partial\omega}{\partial r}
        \frac{\partial\omega}{\partial\theta}e^{-2\lambda}
        e^{-\mu}\text{sin}^2\theta-\frac{1}{Kr^2}
        \frac{\partial\lambda}{\partial\theta}e^{-\mu}
     -\frac{1}{K^2r}\frac{\partial K}{\partial r}
     \frac{\partial\lambda}{\partial\theta}e^{-\mu},    
\end{multline*}
\begin{equation*}
  A(3,0)=-\frac{1}{2}Kr\frac{\partial}{\partial t}\left(
  \frac{\partial\omega}{\partial r}\right)e^{-2\lambda}
   e^{-\mu}\text{sin}\,\theta,
\end{equation*}
\begin{multline*}
   A(3,1)=-\frac{3}{2}K\frac{\partial\omega}{\partial r}
                e^{-\lambda}e^{-2\mu}\text{sin}\,\theta 
       -\frac{1}{2}Kr\frac{\partial^2\omega}{\partial r^2}
                 e^{-\lambda}e^{-2\mu}\text{sin}\,\theta\\
       +\frac{1}{2}Kr\frac{\partial\omega}{\partial r}
             \frac{\partial\lambda}{\partial r}
                 e^{-\lambda}e^{-2\mu}\text{sin}\,\theta
      +\frac{1}{2}Kr\frac{\partial\omega}{\partial r}
           \frac{\partial\mu}{\partial r}e^{-\lambda}
           e^{-2\mu}\text{sin}\,\theta\\
      -\frac{1}{2Kr}\frac{\partial\omega}{\partial\theta}
           \frac{\partial\mu}{\partial\theta}e^{-\lambda}
           \text{sin}\,\theta
      -\frac{3}{2}r\frac{\partial K}{\partial r}\frac
        {\partial\omega}{\partial r}e^{-\lambda}e^{-2\mu}
         \text{sin}\,\theta,
\end{multline*}
\begin{multline*}
   A(3,2)=\frac{1}{2}\frac{\partial\omega}{\partial r}\left(
      \frac{\partial\lambda}{\partial\theta}+\frac{\partial\mu}
      {\partial\theta}\right)e^{-\lambda}e^{-\mu}\text{sin}\,\theta
    -\frac{\partial\omega}{\partial r}e^{-\lambda}e^{-\mu}
         \text{cos}\,\theta\\
     -\frac{1}{K}\frac{\partial K}{\partial\theta}\frac
         {\partial\omega}{\partial r}e^{-\lambda}e^{-\mu}
         \text{sin}\,\theta
     -\frac{1}{2}\frac{\partial^2\omega}{\partial r\partial
         \theta}e^{-\lambda}e^{-\mu}\text{sin}\,\theta.
\end{multline*}
\begin{multline*}
\Omega^0_{\phantom{0}2}=B(1,0)\,\theta^1\wedge\theta^0+B(2,0)\,
    \theta^2\wedge\theta^0+B(3,0)\,\theta^3\wedge\theta^0\\
    +B(3,1)\,\theta^3\wedge\theta^1+B(3,2)\,\theta^3\wedge\theta^2,
\end{multline*}
where $B(1,0)=A(2,0)$,
\begin{multline*}
B(2,0)=\frac{1}{r}\frac{\partial\lambda}{\partial r}e^{-2\mu}
  +\frac{1}{K^2r^2}\frac{\partial^2\lambda}{\partial \theta^2}
  +\frac{1}{K^2r^2}\left(\frac{\partial\lambda}{\partial\theta}
  \right)^2-\frac{1}{K^3r^2}\frac{\partial K}{\partial\theta}
  \frac{\partial\lambda}{\partial\theta}\\
  +\frac{1}{K}\frac
  {\partial K}{\partial r}\frac{\partial\lambda}{\partial r}
  e^{-2\mu}-\frac{3}{4}\left(\frac{\partial\omega}{\partial\theta}
  \right)^2e^{-2\lambda}\text{sin}^2\theta,
\end{multline*}
\[
B(3,0)=-\frac{1}{2}\frac{\partial}{\partial t}\left(\frac
    {\partial\omega}{\partial\theta}\right)e^{-2\lambda}
     \text{sin}\,\theta,
\]
\begin{multline*}
  B(3,1)=-\frac{1}{2r}\frac{\partial\omega}{\partial\theta}e^{-\lambda}
  e^{-\mu}\text{sin}\,\theta -\frac{1}{2}\frac{\partial^2\omega}
  {\partial r\partial\theta}e^{-\lambda}e^{-\mu}\text{sin}\,\theta\\
  +\frac{1}{2}\frac{\partial\omega}{\partial\theta}\frac{\partial
  \lambda}{\partial r}e^{-\lambda}e^{-\mu}\text{sin}\,\theta
  +\frac{1}{2}\frac{\partial\omega}{\partial r}\frac{\partial\mu}
  {\partial\theta}e^{-\lambda}e^{-\mu}\text{sin}\,\theta
  -\frac{1}{2}\frac{\partial\omega}{\partial r}e^{-\lambda}e^{-\mu}
  \text{cos}\,\theta \\ 
  -\frac{1}{2K}\frac{\partial K}{\partial r}\frac{\partial\omega}
   {\partial\theta}e^{-\lambda}e^{-\mu}\text{sin}\,\theta
  -\frac{1}{2K}\frac{\partial K}{\partial\theta}\frac{\partial\omega}
  {\partial r}e^{-\lambda}e^{-\mu}\text{sin}\,\theta,
\end{multline*}
\begin{multline*}
  B(3,2)=-\frac{1}{2Kr}\frac{\partial^2\omega}{\partial\theta^2}
    e^{-\lambda}\text{sin}\,\theta
  +\frac{1}{2Kr}\frac{\partial\omega}{\partial\theta}\frac{\partial
     \lambda}{\partial\theta}e^{-\lambda}\text{sin}\,\theta\\
  -\frac{3}{2Kr}\frac{\partial\omega}{\partial\theta}e^{-\lambda}
      \text{cos}\,\theta 
  -\frac{1}{2}K\frac{\partial\omega}{\partial r}e^{-\lambda}
     e^{-2\mu}\text{sin}\,\theta
  -\frac{1}{2}r\frac{\partial K}{\partial r}\frac{\partial\omega}
   {\partial r}e^{-\lambda}e^{-2\mu}\text{sin}\,\theta\\
  -\frac{1}{K^2r}\frac{\partial K}{\partial\theta}\frac{\partial
   \omega}{\partial\theta}e^{-\lambda}\text{sin}\,\theta.
\end{multline*} 
\begin{equation*}
  \Omega^0_{\phantom{0}3}=C(1,0)\,\theta^1\wedge\theta^0+C(2,0)
   \,\theta^2\wedge\theta^0+C(3,0)\,\theta^3\wedge\theta^0
    +C(2,1)\,\theta^2\wedge\theta^1,
\end{equation*}
where
\[
  C(1,0)=-\frac{1}{2}Kr\frac{\partial}{\partial t}\left(\frac{\partial
  \omega}{\partial r}\right)e^{-2\lambda}e^{-\mu}\text{sin}\,\theta,
\]
\[ 
  C(2,0)=-\frac{1}{2}\frac{\partial}{\partial t}\left(\frac{\partial
  \omega}{\partial\theta}\right)e^{-2\lambda}\text{sin}\,\theta,
\]
\begin{multline*}
  C(3,0)=\frac{1}{r}\frac{\partial\lambda}{\partial r}e^{-2\mu}
  +\frac{1}{K}\frac{\partial K}{\partial r}\frac{\partial\lambda}
   {\partial r}e^{-2\mu}
  +\frac{1}{4}K^2r^2\left(\frac{\partial\omega}{\partial r}\right)^2
  e^{-2\lambda}e^{-2\mu}\text{sin}^2\theta\\
  +\frac{1}{K^2r^2}\frac{\partial\lambda}{\partial\theta}
      \text{cot}\,\theta
   +\frac{1}{K^3r^2}\frac{\partial K}{\partial\theta}
        \frac{\partial\lambda}{\partial\theta}
   +\frac{1}{4}\left(\frac{\partial\omega}{\partial\theta}\right)^2
   e^{-2\lambda}\text{sin}^2\theta,
\end{multline*}   
\begin{multline*}
  C(2,1)=-\frac{1}{2}\frac{\partial\omega}{\partial r}\frac{\partial
    \lambda}{\partial\theta}e^{-\lambda}e^{-\mu}\text{sin}\,\theta
  +\frac{1}{2}\frac{\partial\omega}{\partial\theta}\frac{\partial
    \lambda}{\partial r}e^{-\lambda}e^{-\mu}\text{sin}\,\theta\\
  -\frac{1}{2r}\frac{\partial\omega}{\partial\theta}
     e^{-\lambda}e^{-\mu}\text{sin}\,\theta
  +\frac{1}{2}\frac{\partial\omega}{\partial r}
      e^{-\lambda}e^{-\mu}\text{cos}\,\theta\\
  -\frac{1}{2K}\frac{\partial K}{\partial r}\frac{\partial\omega}
   {\partial\theta}e^{-\lambda}e^{-\mu}\text{sin}\,\theta 
  +\frac{1}{2K}\frac{\partial K}{\partial\theta}\frac{\partial
    \omega}{\partial r}e^{-\lambda}e^{-\mu}\text{sin}\,\theta.  
\end{multline*}
\begin{equation*}
  \Omega^1_{\phantom{0}2}=D(3,0)\,\theta^3\wedge\theta^0
     +D(2,1)\,\theta^2\wedge\theta^1,
\end{equation*}
where
$D(3,0)=-C(2,1)$,
\begin{multline*}
  D(2,1)=\frac{2}{Kr}\frac{\partial K}{\partial r}e^{-2\mu}
  -\frac{1}{r}\frac{\partial\mu}{\partial r}e^{-2\mu}
  +\frac{1}{K^2r^2}\frac{\partial^2\mu}{\partial\theta^2}
  +\frac{1}{K^2r^2}\left(\frac{\partial\mu}{\partial\theta}\right)^2\\
  +\frac{1}{K}\frac{\partial^2 K}{\partial r^2}e^{-2\mu}
  -\frac{1}{K}\frac{\partial K}{\partial r}\frac{\partial\mu}
      {\partial r}e^{-2\mu}
  -\frac{1}{K^3r^2}\frac{\partial K}{\partial\theta}
       \frac{\partial\mu}{\partial\theta}.
\end{multline*}
\begin{multline*}
  \Omega^1_{\phantom{)}3}=E(1,0)\,\theta^1\wedge\theta^0+E(2,0)\,
  \theta^2\wedge\theta^0+E(3,1)\,\theta^3\wedge\theta^1
     +E(3,2)\,\theta^3\wedge\theta^2,
\end{multline*}
where 
$E(1,0)=-A(3,1)$,
$E(2,0)=-B(3,1)$,
\begin{multline*}
  E(3,1)=\frac{2}{Kr}\frac{\partial K}{\partial r}e^{-2\mu}
  -\frac{1}{r}\frac{\partial\mu}{\partial r}e^{-2\mu}
  +\frac{1}{K}\frac{\partial^2 K}{\partial r^2}e^{-2\mu}
  -\frac{1}{K}\frac{\partial K}{\partial r}\frac{\partial\mu}
  {\partial r}e^{-2\mu}\\
  +\frac{1}{4}K^2r^2\left(\frac{\partial\omega}{\partial r}\right)^2
     e^{-2\lambda}e^{-2\mu}\text{sin}^2\theta
  +\frac{1}{K^2r^2}\frac{\partial\mu}{\partial\theta}\text{cot}\,\theta
  +\frac{1}{K^3r^2}\frac{\partial K}{\partial\theta}\frac{\partial
     \mu}{\partial\theta},
\end{multline*}
\begin{multline*}
   E(3,2)=-\frac{1}{Kr^2}\frac{\partial\mu}{\partial\theta}e^{-\mu}
   -\frac{1}{K^2r}\frac{\partial K}{\partial r}\frac{\partial\mu}
       {\partial\theta}e^{-\mu}
   +\frac{1}{K^2r}\frac{\partial^2 K}{\partial r\partial\theta}e^{-\mu}\\
   +\frac{1}{4}Kr\frac{\partial\omega}{\partial r}\frac{\partial
      \omega}{\partial\theta}e^{-2\lambda}e^{-\mu}\text{sin}^2\theta
   -\frac{1}{K^3r}\frac{\partial K}{\partial r}
      \frac{\partial K}{\partial\theta}e^{-\mu}.
\end{multline*}
\begin{multline*}
  \Omega^2_{\phantom{0}3}=F(1,0)\,\theta^1\wedge\theta^0+F(2,0)\,
   \theta^2\wedge\theta^0+F(3,1)\,\theta^3\wedge\theta^1
      +F(3,2)\,\theta^3\wedge\theta^2,
\end{multline*}
where
$F(1,0)=-A(3,2), F(2,0)=-B(3,2), F(3,1)=E(3,2)$,
\begin{multline*}
  F(3,2)=-\frac{1}{K^2r^2}+\frac{1}{r^2}e^{-2\mu}+\frac{2}{Kr}\frac
    {\partial K}{\partial r}e^{-2\mu}
     +\frac{1}{K^2}\left(\frac{\partial K}{\partial r}\right)^2e^{-2\mu}\\
  -\frac{1}{K^4r^2}\left(\frac{\partial K}{\partial\theta}\right)^2
  +\frac{1}{K^3r^2}\frac{\partial^2 K}{\partial\theta^2}
   +\frac{1}{K^3r^2}\frac{\partial K}{\partial\theta}\text{cot}\,\theta
   +\frac{1}{4}\left(\frac{\partial\omega}{\partial\theta}\right)^2
     e^{-2\lambda}\text{sin}^2\theta.
\end{multline*}

\end{document}